\documentclass[12pt]{article}
\begin{document}
\title{\bf Electron in the Einstein-Weyl  space}
\author{S. C. Tiwari \\
\it Institute of Natural Philosophy\\
\it C/o 1 Kusum Kutir, Mahamanapuri\\ 
\it Varanasi 221005, India}
\maketitle
\begin{abstract}
The classical unified theory of Weyl is revisited. The possibility of stable
extended electron model in the Einstein-Weyl space is suggested.
\end{abstract}
\section{Introduction}

The Kaluza-Klein unified theory is well known amongst physicists since the
advent of higher ($> 4$) dimensional theories - specially inspired by
super-string theories, however Weyl's was the first unified theory of
gravitation and electromagnetism \cite{1}. Weyl's theory was rejected on
physical grounds by Einstein, Eddington and Pauli, and Weyl himself abandoned
it later in the light of the developments in quantum theory: thus the
noncompact group of homothetic transformations in the original gauge group was
replaced by the compact circle group of phase transformations. It seems very
few mathematicians have been interested in Weyl geometry, see remarks in
\cite{2}, however Dirac in an attempt to revive this theory in 1973 \cite{3}
noted that, ``Weyl's theory remains as the outstanding one, unrivalled by its
simplicity and beauty''. I believe there are good reasons to explore Weyl
geometry further; (1) influence of Fichtean philosophy on Weyl's thinking
\cite{4} and his idea of continuum that, `a manifold is continuous if the
points are joined together in such a way that it is impossible to single out a
point just for itself, but always only together with a vaguely delimited
surrounding halo, with a neighbourhood' suggest that the study of Weyl
geometry may throw light on some fundamental questions in geometry, and (2)
even if one considers quantum theory to be the fundamental physical theory,
there is a natural generalization of complex Hilbert space of quantum states
incorporating Weyl's vector length holonomy such that the lengths of the
complex vectors in the Hilbert space are allowed to change under parallel
transport, see \cite{5}. Indeed, as Weyl's quantum principle of 1929
\cite{6} \& \cite{7} provided ground work for gauge field theories, it also
suggests the generalization from the invariant scale factor of quantum
mechanics to spaces with a varying scale factor. 

A special case of Weyl's theory, namely Einstein-Weyl manifolds have been
extensively studied, see \cite{2} for a recent review. An Einstein-Weyl space
is defined to be the Weyl space with an extra condition on the curvature,
namely the Einstein-Weyl equation, see Equation(21) below. It is well known
that the question of the inertial mass of the electron and the stability of
the extended structure could not be addressed in the unified theory of Weyl.
In this paper we explore the Einstein-Weyl space to get hints for an electron
model in a classical geometrical framework. Since some of the definitions and
notations are not very familiar in mathematics literature, we summarize them
in the next section. In section 3, the action functional is constructed, and
the field equations are derived using the variational principle. Concluding
remarks constitute the last section.

\section{Rudiments of Weyl geometry}

In the Riemannian geometry, the relative lengths of two vectors at arbitrary
distant points can be compared, therefore for a true infinitesimal geometry it
is natural to enlarge the general coordinate transformations postulating gauge
transformation. Levi-Civita parallelism was used by Weyl to affect this
generalization such that besides a metric
\begin{equation}
ds^2  =  g_{\mu\nu} dx^{\mu}dx^{\nu},				
\end{equation}
there was a linear ground form $A_{\mu}dx^{\mu}$. We confine ourselves to four
dimensions, and adopt the notations and sign-index conventions as given by
Dirac \cite{3} or Eddington \cite{8} with a slight change that $A_{\mu}$ is
used for $k_{\mu}$. Gauge transformation is defined as
\begin{eqnarray}
ds \to ds' &=& \lambda ds \\
A_{\mu} \to A_{\mu}'  &=& A_{\mu} + (\ln \lambda), \mu.		
\end{eqnarray}
For ordinary derivatives comma (,) is used, and covariant derivative is
denoted by Colon (:). A vector under parallel transport from point $x^{\mu}$ to
point $x^{\mu} + dx^{\mu}$ gets a length change given by 	
\begin{equation}
\delta l  =  l A_{\mu} dx^{\mu}.
\end{equation}
Total change in length round a small closed loop is determined by the distance
curvature,
\begin{equation}
F_{\mu\nu}  =  A_{\mu,\nu} - A_{\nu,\mu}.
\end{equation}
Generalized tensors in Weyl space are denoted by $^*T$. Tensors which get
multiplied by $\lambda^n$ under the gauge change (2) are called co-tensors of
power $n$, and in-tensors are gauge-invariant $(n = 0)$. It can be seen that
$g_{\mu\nu}$ is a co-tensor of power 2 and Ö$\sqrt{-g}$ has power 4. One can
define co-covariant or Weyl derivative for any co-tensor. To give an example,
let S be a co-scalar of power $n$ then its co-covariant derivative is 	
\begin{equation}
S_{,^*\mu}  = S_{,\mu}  - n A_{\mu}S.
\end{equation}
A gauge-invariant affine connection can be obtained to be as
\begin{equation}
^*\tau_{\mu\nu}^{\alpha} = \tau_{\mu\nu}^{\alpha} - g_{\mu}^{\alpha}A_{\nu}
- g_{\nu}^{\alpha} A_{\mu}  + g_{\mu\nu}A^{\alpha}. 
\end{equation}
Finally we give the expressions for the generalized Ricci tensor and scalar
curvature
\begin{equation}
^*R_{\mu\nu}  = R_{\mu\nu} - 2F_{\mu\nu} - (A_{\mu,\nu} + A_{\nu,\mu}) - 
g_{\mu\nu}{A^{\alpha}}_{:\alpha} - 2A_{\mu}A_{\nu} +
2g_{\mu\nu}A_{\alpha}A^{\alpha}  
\end{equation}
and
\begin{equation}
^*R = R - {6A^{\mu}}_{:\mu} + 6A_{\mu}A^{\mu}.
\end{equation}
Unlike $R_{\mu\nu}$ which is symmetric in $\mu,\; \nu$ in the Riemann space,
the corresponding tensor $^*R_{\mu\nu}$ contains an antisymmetric part in
the Weyl space. Dirac, in his paper \cite{3}, makes it symmetric by adding few
terms, while Eddington \cite{8} retains the complete expression given above
i.e. Equation (8). It seems the difference between the symmetric and
antisymmetric Ricci tensors in Weyl space may be related to the two kinds of
affine connection \cite{9}. Recall that the covariant derivatives of covariant
and contravariant vectors in the Riemann space are different \cite{8}.
Proceeding in the same way, the generalised affine connections obtained from
the co-covariant derivatives of the covariant and contravariant vectors are
different due to the difference in the Weyl powers. One of them leads to the
symmetric Ricci tensor, while the other one gives Equation (8).

\section{Derivation of Einstein-Weyl equations}

Let us return to the main objective of this paper. Following our earlier work
\cite{10} we present an action functional demanding that the action must be
in-invariant. Let us write the action integral
\begin{equation}
S  =  \int W \sqrt{-g} d^4x.
\end{equation}
Since Ö$\sqrt{-g}$ is a co-invariant of power 4, W must be a co-scalar of power
-4. Noting that $^*R$ is a co-scalar of power -2, one may assume $W = {^*R}^2$;
this was the choice of Weyl \cite{1}. However, this does not give the
Einstein-Weyl equation. A natural assumption preserving the simplicity of the
action for Einstein's field equation in Rieamann space is that $W$ is
proportional to $^*R$. Thus introducing a co-scalar $\chi$ of power -2, it is
proposed that 
\begin{equation}
W =  \xi ^*R +2 C \xi^2.						
\end{equation}
Here $C$ is an arbitrary real number. To derive the Euler-Lagrange equations
of motion, we perform infinitesimal variations in $g_{\mu\nu}$ and $A_{\mu}$ in
the action 
\begin{equation}
S_{E-W} = \int (\xi ^*R + 2C\xi^2)\sqrt{-g} d^4x
\end{equation}
and set $\delta S_{E-W}$ equal to zero (remember that the variations vanish
at and near the boundary of the space-time region). Straightforward
calculations give 
\begin{equation}
\xi \delta (RÖ\sqrt{-g})  = [ \xi (\frac{1}{2} g_{\mu\nu} R - R_{\mu\nu}) +
g^{\mu\nu}{\xi^{:\alpha}}_{\alpha} - \xi^{:\mu:\nu}] \sqrt{-g} \delta
g_{\mu\nu,} 
\end{equation}
\begin{equation}
\xi \delta({A^{\alpha}}_{:\alpha}\sqrt{-g}) = (\xi^{:\mu}A^{\nu} -
\frac{1}{2} \xi_{,\alpha} A^{\alpha} g_{\mu\nu}) \sqrt{-g}\delta g_{\mu\nu} -
\xi^{:\mu} \sqrt{-g} \delta A_{\mu}
\end{equation} 
and
\begin{equation}
\xi\delta(A_{\alpha} A^{\alpha}\sqrt{-g}) = \xi(-A^{\mu} A^{\nu} +
\frac{1}{2} A_{\alpha} A^{\alpha} g^{\mu\nu})\sqrt{-g} \delta g_{\mu\nu} +
2\xi A^{\mu}\sqrt{-g} \delta A_{\mu}. 
\end{equation}
Substituting $^*R$ from Equation (9) in Equation (12) and using the variational
principle we get the following field equations
\begin{eqnarray}
&& C\xi^2 g^{\mu\nu} + \xi(\frac{1}{2}g^{\mu\nu}R - R^{\mu\nu}) +
g^{\mu\nu}{\xi^{:\alpha}}_{\alpha} - 6\xi^{:\mu} A^{\nu} \nonumber \\
&& + 3\xi_{,\alpha} A^{\alpha} g^{\mu\nu} - \xi^{:\mu:\nu} - 
6\xi A^{\mu} A^{\nu} + 3\xi A_{\alpha}A^{\alpha} g^{\mu\nu}  =  0
\end{eqnarray}
and  
\begin{equation}
\xi^{:\mu} + 2\xi A^{\mu} = 0.		
\end{equation}
Using Equation (17), we can rewrite Equation (16) in the form
\begin{equation}
R^{\mu\nu} = g^{\mu\nu}\Lambda + 2A^{\mu}A^{\nu} + 2A^{\mu:\nu}, 
\end{equation}
where
\begin{equation}
\Lambda = C\xi + \frac{1}{2}R + A_{\alpha}A^{\alpha} - {2A^{\alpha}}_{:\alpha}.
\end{equation}
Variation of $\xi$ in the action, Equation (12), gives
\begin{equation}
\xi = - \frac{^*R}{4C}.						
\end{equation}
Equation (20) is not an independent one; contraction of Equation (18) gives
Equation (20). Substituting $\xi$ in Equation (18), and rewriting the term
$2A^{\mu:\nu}$ in symmetric and skew-symmetric parts, we obtain the
Einstein-Weyl equation 
\begin{equation}
^*S^{\mu\nu} = \frac{1}{4} {^*R} g^{\mu\nu}.
\end{equation}
Here $^*S^{\mu\nu}$ is the symmetric part of the Ricci tensor
\begin{equation}
^*R^{\mu\nu} = {^*S}^{\mu\nu} - 2F^{\mu\nu}.
\end{equation}
Evidently
\begin{equation}
F^{\mu\nu} =  0							
\end{equation}
in view of the rearranged form of $2A^{\mu:\nu}$ used in Equation (18)
\begin{equation}
2A^{\mu:\nu}  =  A^{\mu:\nu} + A^{\nu:\mu} - F^{\mu\nu}.		
\end{equation}
From Equation (23) it is clear that $A_{\mu}$ is locally a gradient. This
result is consistent with the contracted Bianchi identity for a Weyl space
\cite{11}, see also \cite{12}. In general
\begin{equation}
{(^*S^{\mu\nu} - \frac{1}{2}{^*R} g^{\mu\nu})}_{:^*\nu} + {F^{\mu\nu}}_{:
^*\nu} = 0. \end{equation}
In 4 dimensions, the Weyl derivative of $F^{\mu\nu}$ is an ordinary covariant
divergence, and for the E-W space use can be made of Equation (21) to reduce
the Equation (25) in to the form
\begin{equation}
{F^{\mu\nu}}_{:\nu} = \frac{1}{4}({^*R}^{:\mu} + 2{^*R}A^{\mu}).
\end{equation}
Substituting Equation (20) and Equation (23) in Equation (26), we get
back Equation (17). 	

This derivation leads to the Einstein-Weyl equations, but for a restricted
class of locally conformal Einstein spaces. Examples are known, see \cite{2},
such that $F_{\mu\nu}$, is non-vanishing. E-W Equations (21) do not contain an
energy-momentum tensor for the field $F_{\mu\nu}$. Therefore this result,
namely that the natural E-W space to be a locally conformal Einstein space,
seems significant. Is it possible to have a non-zero $F_{\mu\nu}$ together
with the E-W equations ? The Bianchi identity Equation (26) indicates that
r.h.s. of this equation can be interpreted as a source current density for the
Maxwell field. Thus, Equation (17) can be generalized to
\begin{equation}
\xi^{:\mu} + 2\xi A^{\mu}  = J^{\mu}.
\end{equation}
For a  consistent derivation of the field equations, the action function,
Equation (11) should be changed to
\begin{equation}
W  = \xi {^*R} + 2\xi^2 + pF^{\mu\nu} F_{\mu\nu}.	
\end{equation}
Here $p$ is an arbitrary real number, and $C$ is set equal to 1. Variation in
$g_{\mu\nu}$ contributes an additional term $2pE^{\mu\nu}$, with the
electromagnetic energy-momentum tensor being
\begin{equation}
E^{\mu\nu} = \frac{1}{4} g^{\mu\nu} F^{\lambda\sigma} F_{\lambda\sigma}
- F^{\mu\lambda}{F^{\nu}}_{\lambda}. 
\end{equation}
Varying $A_{\mu}$, instead of Equation (17) we get
\begin{equation}
{F^{\mu\nu}}_{:\nu} = \frac{3}{2p}(\xi^{:\mu} + 2\xi A^{\mu}).
\end{equation}
In order to identify Equation (30) with the Bianchi identity, Equation (26),
for the E-W space, we assume $p = -3/2$. The co-scalar field $\xi$ determines
the current density 4-vector, and Equation (27) can be used to arrive at the
following equations:
\begin{equation}
\xi^2 g^{\mu\nu} + \frac{1}{2}\xi {^*R} g^{\mu\nu} - \xi {^*S}^{\mu\nu} + B1 +
B2 =  0, 
\end{equation}
\begin{equation}
B1  = \xi F^{\mu\nu} - \frac{1}{2}[J^{\mu:\nu} - J^{\nu::\mu}] - 
4[J^{\mu}A^{\nu} - J^{\nu} A^{\mu}]
\end{equation}
and
\begin{equation}
B2  = -3E^{\mu\nu} - \frac{1}{2}[J^{\mu:\nu} + J^{\nu:\mu}] - 
2 [J^{\mu}A^{\nu} + J^{\nu}A^{\mu}] + J_{\alpha}A^{\alpha}g^{\mu\nu}. 
\end{equation}
Since $B1$ and $B2$ are trace-less, contraction of Equation (31) gives
Equation (20) with $C = 1$. The skew-symmetric part, $B1$, must vanish. It is
further postulated that the symmetric part $B2$ is also zero. Equation (31)
then reduces to the E-W Equation (21). 	

At first, assuming $B1$ and $B2$ to be zero may look arbitrary, but returning
to the fundamental unresolved problem of the stability of a purely
electromagnetic model of electron we gain a new insight for analysing Equation
(32) and Equation (33). To simplify the problem, let us consider the E-W space
for $\xi = $ a constant. The current continuity equation gives the Lorentz
gauge condition as 
\begin{equation}
{A^{\mu}}_{:\mu}  =  0,							
\end{equation}
and Equation (20) becomes
\begin{equation}
R + 6A_{\mu} A^{\mu} =  {\rm constant}.				
\end{equation}
Since $J^{\mu} = 2\xi A^{\mu}$, $B1$ is identically zero. Equation (33) shows
that for $B2$ to be zero, the electromagnetic energy momentum tensor must
balance with a stress tensor of charge density in which the constant $\xi$
appears. The implication that $B2 = 0$ is reminiscent of the Poincare stress
for an electron model \cite{13} and the problem of inertial mass and
electromagnetic energy in a Weyl space discussed by Weyl \cite{1} and Eddington
\cite{8}. 

\section{Conclusion}

A natural Einstein-Weyl space having pure gauge fields $A_{\mu}$ is shown to
follow from a simple action principle. A non-vanishing distance curvature (or
the Maxwell field) and the contracted Bianchi identity in the Weyl space
indicate a non-trivial generalisation postulating current-density four-vector.
The general field Equation (31), is shown to admit a special class of space,
namely the E-W space if certain conditions (i.e. $B1 = B2 = 0$) are imposed.
Physical arguments emanating from the stability of the extended model of the
electron seem to justify these conditions. However, construction of an
actual physical model of the electron remains as an exciting possibility. We
conclude this paper by asking: "Is the interior of electron an  E-W
space?

The library facility of the Banaras Hindu University is acknowledged. I am
grateful to Prof. H. Pedersen and Prof. P. Goudochon for useful discussion on
the E-W spaces. I thank the anonymous reviewers for constructive comments.

\end{document}